\documentclass[twocolumn,prb,floats,showpacs,superscriptaddress]{revtex4}
\usepackage{graphicx,epsfig}
\usepackage{times}
\usepackage{graphics,dcolumn,bm,float}
\usepackage{amssymb,amsmath,rotate,color}
\usepackage{latexsym}

\begin{document}
\title{Generation of fully spin-polarized currents in three-terminal graphene-based transistors }
\author{R. Farghadan}\email{rfarghadan@kashanu.ac.ir}
\affiliation{Department of Physics, University of Kashan, Kashan, Iran }
\author{A. Saffarzadeh}
\affiliation{Department of Physics, Payame Noor University, P.O.
Box 19395-3697 Tehran, Iran} \affiliation{Department of Physics,
Simon Fraser University, Burnaby, British Columbia, Canada V5A
1S6}
\date{\today}

\begin{abstract}
We propose three-terminal spin devices with graphene nanoribbons
(terminals) and a graphene flake (channel) to generate a highly
spin-polarized current without an external magnetic field or
ferromagnetic electrodes. The Hubbard repulsion within the
mean-field approximation plays the main role to separate the
unpolarized electric current at the source terminal into
spin-polarized currents at the drain terminals. It is shown that
by modulating one of the drain voltages, a fully spin-polarized
current can be generated in the other drain terminal. In addition,
the geometry of the channel and the arrangement of edge atoms have
significant impact on the efficiency of spin currents in the
three-terminal junctions which might be utilized in generation of
graphene-based spin transistors.
\end{abstract}
\maketitle

\section{Introduction}

The ability to control electron transport by an external electric
field and fabrication of well-defined graphene structure has
attracted much experimental and theoretical interest in
nanoelectronics \cite{Perez,Jia}.
Graphene-based transistors are considered as promising candidates
for post-silicon electronics
\cite{Schwierz,Tombros,Liao,Peres,Lin} due to their high carrier
mobility, high field transport, short gate possibility and band
gap engineering. Graphene might also be a promising material for
fabrication of spintronic devices due to long spin-relaxation time
and length, the low intrinsic spin-orbit coupling and the
hyperfine interactions \cite{Tombros,Kim}. For instance, a large
magnetoresistance was observed in graphene nanoribbon field-effect
transistors by applying a perpendicular magnetic field\cite{Bai}.

Most of the previous studies in spin field-effect transistors
demonstrate spin filtering effect by utilizing graphene and/or
organic molecules in contact with magnetic electrodes
\cite{Rocha,Sugahara,Saffarzadeh,Zhang}. Interestingly, it has
been shown that zigzag-shaped graphene nanoribbons and quantum
dots have spin-polarized features both theoretically and
experimentally \cite{Joly,Rossier,Fujita}. Moreover,
spin-polarized currents induced by zigzag-edge states of
graphene-based structures with various geometries were intensively
investigated \cite{Guo,Rojas,Farghadan}. However, many
details in developing and engineering field-effect transistor
based on graphene is still unclear. For instance, the high contact
resistance between graphene and metallic source, drain, and gate
electrodes are among important challenges in developing transistor
designs \cite{Russo,Schwierz}. On the other hand, in the case of
spintronic devices, it is very important to find nonmagnetic
materials and design appropriate nano-structures where a
spin-polarized current can be injected and flowed without becoming
depolarized\cite{SaffrzadehR}. Historically, various
three-terminal structures for generating spin currents with
spin-orbit interaction, ferromagnetic contacts and interference
effect have been proposed \cite{Saha,Pareek,LF,BA,Chi,LWang,Dey}.

Here, we propose novel spin-dependent Y-shaped and T-shaped
transistors based solely on carbon atoms and intrinsic magnetism
in zigzag edges for generating fully spin-polarized currents in
the system. We utilize armchair and zigzag graphene nanoribbons as
electrodes and zigzag-edge graphene flakes as channels. The effect
of structural difference in the channels and also in the
electrodes on spin-polarized currents will be examined. In these
spin-dependent three-terminal devices which consist of two drain
terminals, one of the drain voltages is adjusted to provide the
required conditions to generate a fully spin-polarized current
into the second drain terminal. The induced localized magnetic
moments on the zigzag edges of the channel are the main source in
generating spin currents between source and drain electrodes. It
is shown that the spectrum of the spin-polarized current which is
produced without utilizing a magnetic field or magnetic materials, depends
on the geometry of the channel, carbon-based electrodes, and also
the atomic arrangement of the edges.

\section{Model and Method}
We first consider the spin transistor effect in a three-terminal
Y-shaped junction as shown in Fig. 1. The device consists of a
hexagonal graphene flake (channel) sandwiched between three
perfect semi-infinite armchair graphene nanoribbons. The total
Hamiltonian is described by the $\pi$ orbital tight-binding model
including the Hubbard repulsion with the mean-field approximation
in which the electron-electron interaction induces localized
magnetic moments on the zigzag-shaped edges. The mean-field
Hamiltonian for the graphene flak can be written as
\cite{Farghadan,Guo}:
\begin{equation}
H_{C}= t\sum_{<i,j>,\sigma} c^{\dagger}_{i,\sigma}c_{j,\sigma}
+U\sum_{i,\sigma}\hat{n}_{i,\sigma}\langle
\hat{n}_{i,-\sigma}\rangle \ ,
\end{equation}
where the operator $c^{\dagger}_{i\sigma}(c_{i\sigma})$ creates
(annihilates) an electron with spin $\sigma$ at site $i$ and
$n_{i\sigma}=c^{\dagger}_{i\sigma}c_{i\sigma}$ is a number
operator.  The first term corresponds to the single $\pi$-orbital
tight-binding Hamiltonian with hopping parameter $t = -2.66$ eV,
while the second term accounts for the on-site Coulomb interaction
with $U = 2.82$ eV. To study the electronic and spin transport in
this model, we use the non-equilibrium Green's function method in
which the Green's function of the channel is expressed as
\begin{equation}
\hat{G}_{C}(\varepsilon)=[\varepsilon\hat{I}-\hat{H}_{C}-\sum_{q=1}^3\hat\Sigma_{q}(\varepsilon)]^{-1},
\end{equation}
where $\hat{\Sigma}_{q}(\varepsilon)$ is the self-energy matrix
which contains the influence of electronic structure of the $q$th
graphene nanoribbon (terminal) through the surface
Green's function \cite{Sancho}. Here, one of the leads is
considered as the source terminal (Source) and the other leads
serve as drain terminals (Drain 1 and Drain 2). The magnetic
moment (spin) on each atomic site of the channel is expressed as
$m_i= \langle {S_i}\rangle=(\langle
\hat{n}_{i,\uparrow}\rangle-\langle
\hat{n}_{i,\downarrow}\rangle)/2$ where $\hat{n}_{i,\sigma}$ is
the number operator at $i$th site. Thus, by neglecting spin-fillip scattering,
the transmission probability, $T_{p,q}^\sigma$, for electrons transmitted from
terminal $p$ to terminal $q$ in spin channel $\sigma$ ($=\uparrow$ or $\downarrow$)
can be written as $ T_{p,q}^\sigma(\varepsilon)=\mathrm{Tr}[\hat{\Gamma}_{p}
\hat{G}_{C}\hat{\Gamma}_{q}\hat{G}_{C}^{\dagger}]^\sigma\  ,
$ where the coupling matrices $\hat\Gamma_p$ are expressed as
$\hat\Gamma_p=-2\,\mathrm{Im}[\hat\Sigma_{p}(\varepsilon)]$
\cite{Datta}. It is clear that the transmission probabilities are
only spin and energy dependent.
\begin{figure}
\centerline{\includegraphics[width=0.75\linewidth]{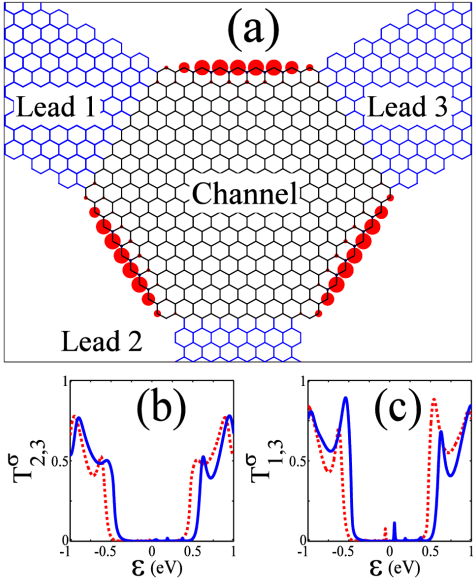}}
\caption{(Color online) (a) Schematic view of the three-terminal
Y-shaped junction with localized magnetic moments on zigzag
edges. The red circles correspond to the majority spin electrons.
The lead 1 is the source terminal and the leads 2 and 3 are the
drain terminals. Spin-dependent transmission coefficients (b)
between leads 2 and 3, and (c) between leads 1 and 3 as a function
of electron energy. The solid (blue) and dashed (red) curves
correspond to spin-up and spin-down electrons, respectively.}
\end{figure}

The net  spin current $\sigma$,
$I_p^\sigma$, flowing into terminal $p$ is obtained through the
multi-terminal Landauder- B\"{u}ttiker formula
\cite{Pareek,Datta},
\begin{equation}
I_p^\sigma=e^2/h \sum _q(T_{p,q}^\sigma V_p-T_{q,p}^\sigma V_q),
\end{equation}
where $V_p$ is the voltage at terminal $p$. Therefore, the net
charge current flowing through terminal $p$ is expressed as
$I_p=I_p^\uparrow+I_p^\downarrow$. To obtain a fully
spin-polarized current, one of the conducting spin channels must
be blocked. For instance, to generate a net spin current $\sigma$
into lead 3 of Fig. 1, we set $I_3^{\bar{\sigma}}=0$ in Eq. 3. If
we consider lead 1 as the source terminal with voltage
$V_1=0$, and leads 2 and 3 as drain terminals with voltages $V_2$,
and $V_3=V_0$, respectively, the required voltage $V_2$ in terminal 2 for blocking
electrons with spin $\bar{\sigma}$ in terminal 3 can be obtained as
\begin{equation}\label{V}
V_2^{\bar{\sigma}}/V_0=
[T_{3,1}^{\bar{\sigma}}(\varepsilon)+T_{3,2}^{\bar{\sigma}}(\varepsilon)]/T_{2,3}^{\bar{\sigma}}(\varepsilon)\
,
\end{equation}
which is applied to both spin-up and spin-down electrons. The superscript
$\bar{\sigma}$ in $V_2^{\bar{\sigma}}$ is used to indicate that only the
net spin current $\bar{\sigma}$ flowing through the terminal 3 is blocked and,
thus, the net charge current can be written as
\begin{equation}
I_3=I_3^{\sigma}=\frac{e^2V_0}{h}\left[\left(\frac{V_2^{\bar{\sigma}}}{V_0}
\right)T_{3,2}^\sigma-(T_{1,3}^\sigma+T_{2,3}^\sigma)\right]\  .
\end{equation}

Therefore, by tuning the voltage in one of the drain terminals,
a net spin current can be achieved in the other drain terminal. This is a novel mechanism to
produce a fully spin-polarized current in graphene-based
transistors in the absence of spin-orbit interaction.

\begin{figure}
\centerline{\includegraphics[width=0.7\linewidth]{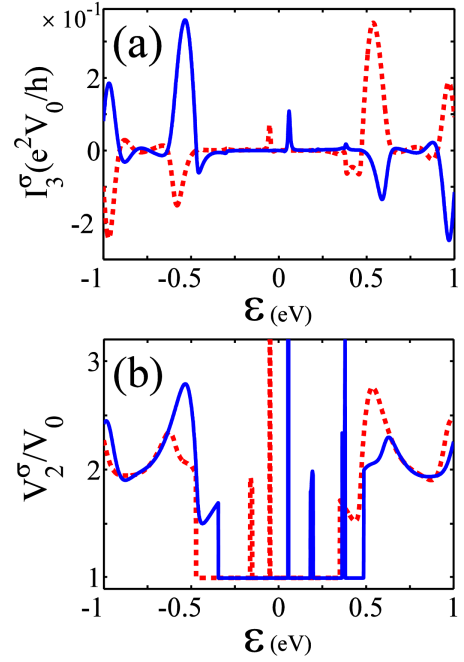}}
\caption{(Color online) (a) The fully spin-polarized currents in terminal 3
and (b) the ratio of voltages $V_2^\sigma/V_0$ as a function of electron
energy. The solid (blue) and dashed (red) curves correspond to
spin-up and spin-down electrons, respectively.}
\end{figure}

Note that since we deal with the ratio of drain voltages, to keep
the transport in ballistic regime, both the voltage values $V_0$
and $V_2^\sigma$ are assumed to be very small which imply that
the electrons are transmitted through the channel with the energies
close to the Fermi level ($E_F=$0 eV). It is evident from Eq. (\ref{V})
that the voltage $V_2^{\sigma}$ is energy dependent through the transmission
probabilities $T_{p,q}^\sigma(\varepsilon)$. As a result, the spin
current $I^\sigma_p$ is a function of both the electron energy
$\varepsilon$ and the drain voltage $V_2^\sigma$, as it is discussed below.
Moreover, in Eq. (3) we have assumed that the applied biases are very small
and, hence, $T_{p,q}^\sigma$ are unaffected by the bias voltages. This is
reasonable, because an all-graphene three-terminal system, like the ones we
study here, can be considered as a uniform system with negligible contact
resistance \cite{Topsakal}.

\begin{figure}
\centerline{\includegraphics[width=0.75\linewidth]{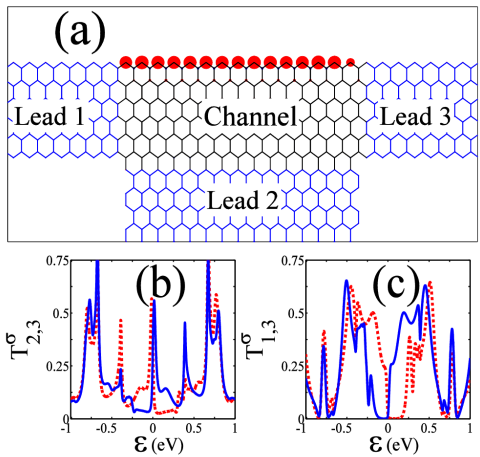}}
\caption{(a) Schematic view of a three-terminal T-shaped
junction with localized magnetic moments at zigzag edge; the red
circles correspond to the majority spin electrons. (b) and (c) show
the spin- and energy-dependent transmission coefficient between the drain 1 and drain 2 and between source
and drain 2, respectively.  The solid (blue) and dashed (red) curves
correspond to spin-up and spin-down electrons.}
\end{figure}

\section{Results and discussion}
Now, we present the numerical results of spin-dependent
transmission probabilities and net spin currents in three-terminal
junctions with two different geometries: hexagonal (Y-shaped) and
rectangular (T-shaped) structures. As shown in Fig. 1(a), in the
Y-shaped junction a hexagonal-shape graphene flake (channel) with
zigzag edges is sandwiched between three perfect semi-infinite
armchair graphene nanoribbons (electrodes). The hexagonal channel
consisting of 486 carbon atoms has the same number of A- and
B-type atoms and according to Lieb's theorem \cite{Lieb}, the
total magnetic moment of the ground state in the graphene flake is
zero. However, by choosing an appropriate design for the electrodes
in a way that the semi-infinite armchair nanoribbons are connected to only
one type of carbon atoms of the channel, one can induce an imbalance in the
edge magnetism between the two sublattices A and B. In this context, the
total magnetic moment of the channel in Fig. 1(a) increases from
zero to $3.7\mu_B$ and, hence, a spin current without magnetic
electrodes is produced. In the proposed structure the drain
terminals have different widths, indicating that the rotational
symmetry in the junction is broken and the transmission
coefficients and spin currents in drains 1 and 2 are different.

\begin{figure}
\centerline{\includegraphics[width=0.7\linewidth]{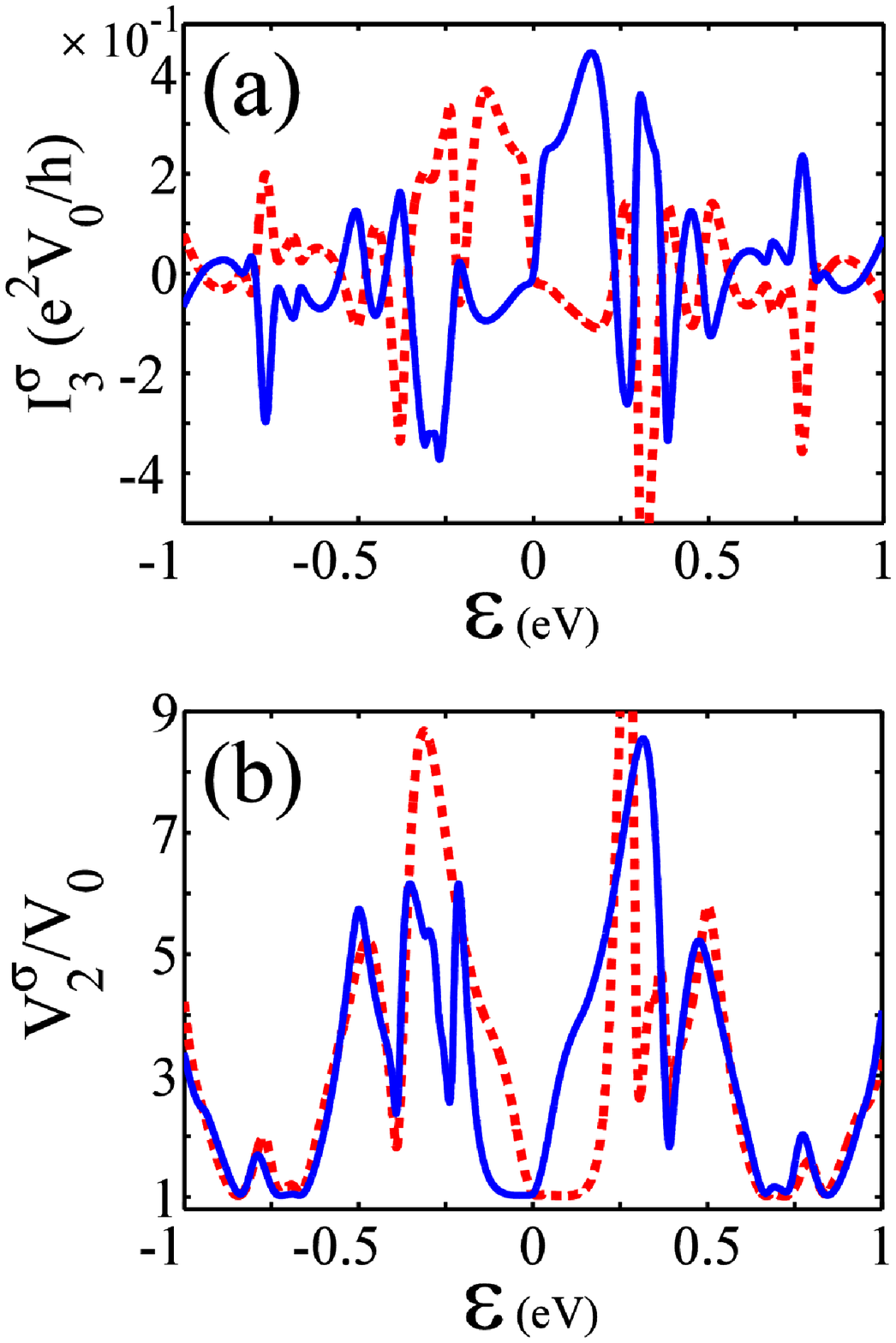}}
\caption{(Color online) (a) The fully spin-polarized currents in
terminal 3 and (b) the ratio of voltages
$V_2^\sigma(\varepsilon)/V_0$ as a function of electron energy. The
solid (blue) and dashed (red) curves correspond to spin-up and
spin-down electrons, respectively. }
\end{figure}

The generation mechanism of the fully spin-polarized current
relies solely on the magnetic edge states in the channel. When an
unpolarized charge current flows in the channel, due to its
interaction with the localized magnetic moments, it becomes
spin-polarized and, hence, two spin-up and spin down currents with
different transmission probabilities are generated. As a result,
by using Eq. 4, we can find an appropriate electrostatic potential
for electrons with energy $\varepsilon$ in terminal 2 to block one
of the spin currents in terminal 3. This implies that the charge current
flowing in terminal 3 (see Eq. 5) is a net spin (-up or -down)
current, while the both spin-up and spin-down currents in terminal
2 remain nonzero.

The spin-dependent transmission coefficients from terminal 2 to
terminal 3, $T^\sigma_{2,3}$, and from terminal 1 to terminal 3,
$T^\sigma_{1,3}$, as a function of electron energy, $\varepsilon$,
are depicted in Figs. 1(b) and 1(c), respectively. It is shown that
the transmission spectra for the two spin subbands are
approximately split in the all electronic states within the shown
energy windows. This feature implies that the three-terminal
graphene transistor at ground state can be utilized for generation
of spin-polarized currents without including spin-orbit
interaction or applying magnetic fields. In other words, the
separation between the two spin subbands which provide two paths
for electron conduction through the junction well explains the
spin-filter effect in this type of transistors. Since nonmagnetic
electrodes are used in this structure, it is evident that the
emergence of magnetic properties is due to the edge localized
magnetic moments of the hexagonal graphene flake. Moreover, the
spin-dependent transmission coefficients show various spectra
around the Fermi energy, depending on the electrode thickness, so
that in this structure the thicker electrode transmits electrons
slightly more than the thinner one (not shown here). The
transmission coefficients $T_{1,2}$ and $T_{3,2}$ have the same
value due to the mirror symmetry in the junction. Note that, the
transmission coefficients obey the relation $\sum_p
T_{p,q}^\sigma= \sum_p T_{q,p}^\sigma$ at equilibrium
\cite{Datta}.

In order to achieve a fully spin-polarized current, for an
electron with energy $\varepsilon$ and spin $\sigma$ the drain
voltage $V_2^\sigma(\varepsilon)$ is tuned through the constrain
imposed by Eq. 4 which allows us to control the spin polarization
of electrical current in the lead 3. The net spin current
$I_3^\sigma$ as a function of electron energy is shown in Fig.
2(a) when the drain voltage $V_2^\sigma$ for the same electron energies
takes the values shown in Fig. 2(b). We see that
the current in lead 3 for the some energy ranges is fully spin
polarized. In fact, for all electron energies at which
$V_2^\uparrow(\varepsilon)$ and $V_2^\downarrow(\varepsilon)$ of
Fig. 2(b) do not take the same values, a fully spin-polarized
current in lead 3 is generated. In other words, if the voltage
in terminal 2 takes the value $V_2^\uparrow(\neq
V_2^\downarrow)$, only $I_3^\downarrow$ will flow into the lead 3,
while in the case of $V_2^\uparrow=V_2^\downarrow$ both
current curves in Fig. 2(a) cross each other and this feature
happens only if $I_3^\uparrow=I_3^\downarrow=0$. Therefore, the
dependence of spin currents in lead 3 on the drain voltage $V_2$
reveals a voltage-controlled spin transport in this type of
Y-shaped transistors.

In Fig. 3(a) we consider a T-shaped junction consisting of a
rectangular graphene flake (channel), sandwiched between two
perfect semi-infinite zigzag nanoribbons as source (lead 1) and
drain 2 (lead 3) terminals, and one semi-infinite armchair
nanoribbon as drain 1 (lead 2) terminal. The rectangular channel
with 180 carbon atoms has the same number of A- and B-type atoms
and hence $\sum_im_i=0$. By including the effect of semi-infinite
electrodes as shown in Fig. 3(a), however, the total magnetic
moment increases to $1.8 \mu_B$. Density functional theory
calculations predict that the spin-correlation length limits the
long-range magnetic order to 1 nm at room temperature
\cite{Yazyev}. Therefore, the electron-electron interaction is
limited in the channel region. In other words, the effect of
induced magnetic moments on the zigzag-shaped edges in the source
and drain terminals are not intentionally included in our
calculations to emphasize on the role of channel geometry and its
magnetic edge states in generating spin-polarized currents in such
graphene nano-transistors without magnetic electrodes.

Comparing the transmission coefficients shown in Fig. 3(b) and
3(c) it is clear that due to the influence of edge states in the
channel, the spectra are spin polarized and dependent on the
electrode geometries. In contrast to the hexagonal junction shown
in Fig. 1, the transmission coefficient of the rectangular
junction does not vanish at the band center due to the
zigzag-shaped edges in the source and drain terminals. The
spin-dependent transmissions in such three-terminal junctions are
sensitive to the width of the electrodes. The drain terminal 2 in
Fig. 3(a) is wider than the other terminals. The width of terminal 2 also
affects the magnitude of the induced magnetism and, hence, the
generation of net spin currents becomes considerably size-dependent.

Figure 4 shows the spin-polarized electric currents in the terminal 3
and the ratio of voltages $V_2^\sigma(\varepsilon)/V_0$ as a
function of energy. The result reveals that the spin-dependent
electron transport can be controlled by modulating the drain
voltage 2. In addition, it is possible to alter the magnitude and
the sign of spin currents if an appropriate drain voltage is
chosen. In this way, the spin-dependent currents can be easily
reversed and tuned between the source and drain terminals without
requiring magnetic fields or electrodes, which may open a way for
generation of voltage-driven spintronic devices \cite{SciRep}.

Comparing Fig. 2 with Fig. 4, we find that the T-shaped structure
at energies around Fermi level can function as a spin current switch
in a more efficient way than the Y-shaped junction. This feature comes
from zero-energy edge states \cite{Wakabayashi} in the zigzag graphene
nanoribbons (leads) 1 and 3 of Fig. 3(a) which act as the source and the
drain terminals, respectively. In fact, at very low bias, only electrons
with energies around Fermi level acquire a chance to travel from the
source to the drain. Therefore, by tuning the voltage in the terminal 2,
one can obtain a fully spin-polarized current (see Fig. 4(a)). Due to
the absence of zero-energy edge states in the armchair graphene
nanoribbons (leads) of the Y-shaped structure, a gap opens in the
spin currents which vanishes the switching behavior around zero energy.
Moreover, the geometry of the edges in the proposed T-shaped and Y-shaped
junctions shows that the zigzag edges in the channel play the main role
in the process of net spin current generation and therefore, the chirality
of the leads, whether zigzag or armchair, is not crucial in the process.
We should also mention that since the value of induced total magnetic moment
in the Y-shaped and T-shaped junctions presented here is maximum among other
three-terminal structures (not shown), it is believed that the two structures
are optimal.

Note that in most three-terminal systems the spin-orbit
interaction which couples the spin of electrons with their motion
is the main source in generation of pure spin currents \cite{Lu,
Chi,LWang,Dey}. Moreover, the spin-orbit interaction may lead to
spin-flip scattering between electronic states with opposite spin
orientations which lessens the degree of spin polarization in
multi-terminal structures. In the carbon-based devices, however,
the strength of intrinsic spin-orbit coupling is weak. Therefore,
by designing a graphene-based three-terminal structure and
utilizing the property of magnetic edge states, one can generate a
net spin current without applying a magnetic field or coupling
ferromagnetic leads \cite{LF,Yamamoto,Chi}.

\section{Conclusion}
In conclusion, we have proposed three-terminal spin transistors
based solely on carbon atoms and shown that by modulating voltage
in one of the drain terminals the spin currents in the other drain
terminal will be highly affected, so that a fully spin-polarized
currents can be achieved. We found that the geometry of graphene
flakes, the arrangement of edge atoms and the type of graphene
nanoribbons taken in each terminal have significant impacts on the
voltage required to generate a fully spin-polarized current. The
advantage of such carbon-based spin transistors is utilizing
non-magnetic electrodes which reduces manufacturing cost and the
spin-dependent electron scattering. Moreover, the graphene
nanoribbons are matched more conveniently to the graphene flakes
and reduce the contact resistance. Therefore, the three-terminal
spin devices in which the electron transmission is sensitive to
the magnetic configurations of the localized moments and magnitude
of the drain potentials can be an attractive pathway for designing
spintronic devices with nonmagnetic electrodes.

Although the device fabrication and experimental measurement of
spin currents in the well-defined zigzag edges are important
challenges in this type of junctions, the Y-shaped nanoribbons
\cite{Papon} with zigzag edges and Y-junction carbon nanotubes
\cite{Li} with magnetic impurity may be experimentally
useful for testing fully spin-polarized currents in such spin transistors.

\section*{Acknowledgement}
This work financially supported by Iran National Support Foundation: INSF.

\end{document}